# REPRODUCTION OF HOLOGRAM IMAGE USING A ZONE PLATE FOR HARD X-RAY RADIATION

A. V. Kuyumchyan[1], A. Yu. Souvorov[2], T. Ishikawa[2], A. A. Isoyan[1], V.V. Aristov[1], K. Trouni[3], and E. Sarkisian[3]

[1]Institute of Microelectronics Technology and High Purity Materials of RAS, Russian Federation, Chernogolovka, 142432,
[2]SPring-8, JASRI, 1-1-1 Kouto, Mikazuki-cho, Sayo-gun Hyogo 679 5198, Japan and
[3]International Academy of Science and Technology, P.O. Box 4385, CA 91222, USA

The images of the silicon test object have been studied. An image for in-line hologram for hard X-ray (8 – 18 KeV) is presented. The transmission of a hologram image for hard X-ray radiation using Fresnel phase zone plate has been investigated. The experimental investigations have been conducted on the station BL29XU, Spring-8.

**Introduction**

The development of X-ray optics opens new possibilities for diagnostics in the field of Biology, microelectronics and nanotechnologies. Methods of X-ray holography for soft X-ray radiation [1,2] and transmission of the image of phase objects for both hard and soft X-ray radiation [3-5] have rapidly develop in recent years. The absence of coherent beams (X-ray lasers) and recording devices for three-dimension images is the main difficulty to obtain X-ray holograms by optical method. In the present work the following experimental results are given: recordings of a Gabor hologram for hard X-ray radiation and transmission of a hologram image using a silicon phase zone plate (ZP) [6].

**Experimental scheme and results**

The images of the test silicon object consisting of the word-combination "X-RAY phase contrast, IPMT RAS CHERNOGOLOVKA" have been investigated. The test objects have been prepared on the silicon crystal membranes 4 $\mu$m thick by electron-beam lithography and ion-plasma etching. The width of the small letters is 2 $\mu$m and the height is 3 $\mu$m for one case and 1.5 $\mu$m for the other (Fig.1a). The transmission coefficient is equal to 98.9% for 3 $\mu$m and 99.4% for 1.5 $\mu$m (the X-ray radiation energy is 12.4 KeV).

The experimental investigations have been conducted on the station BL29XU, Spring-8 [7]. Two different experimental schemes are considered: 1) The experimental scheme to obtain a Gabor hologram at the X-ray radiation energy in the range 8-18keV. (Fig.1b). 2) The experimental scheme to transmit a hologram image at the X-ray radiation energy of 10kev (fig.1c).
In the first case an X-ray beam of synchrotron radiation from the 25 $\mu$m source is monochromatized by a two-crystal monochromator Si (111), then it is transmitted to the test object as far as 987.35 m which is set perpendicularly to the falling X-ray beam. Behind the test object an interference image is formed, which is recorded by a CCD-camera with the resolving power 0.3 $\mu$m at the distance from 0.03 m up to 3 m. The images of interference patterns at the different distances from the test object have been examined. The experimental patterns for different distances from the test object at the 12.4 keV energy of the falling X-ray beam are shown in Figures 2 (a,b,c,d). As is seen in the patterns, the image contrast changes depending on the distance. The experimental investigations have shown that the images appear beginning with the particular distance from the test object (e.g. for 12.4 KeV energy this distance is 25 mm). To confirm the reliability of the results obtained the additional experiments have been conducted. The image of the phase test object with the diffuser has been obtained. It is well known that the phase contrast can be extinguished by a diffuser. Experimental investigations have shown that a diffuser insignificantly affects interference images. The contrast has been shown to improve for high energies when changing the energy of the falling X-ray beam from 8 KeV to 18 KeV.

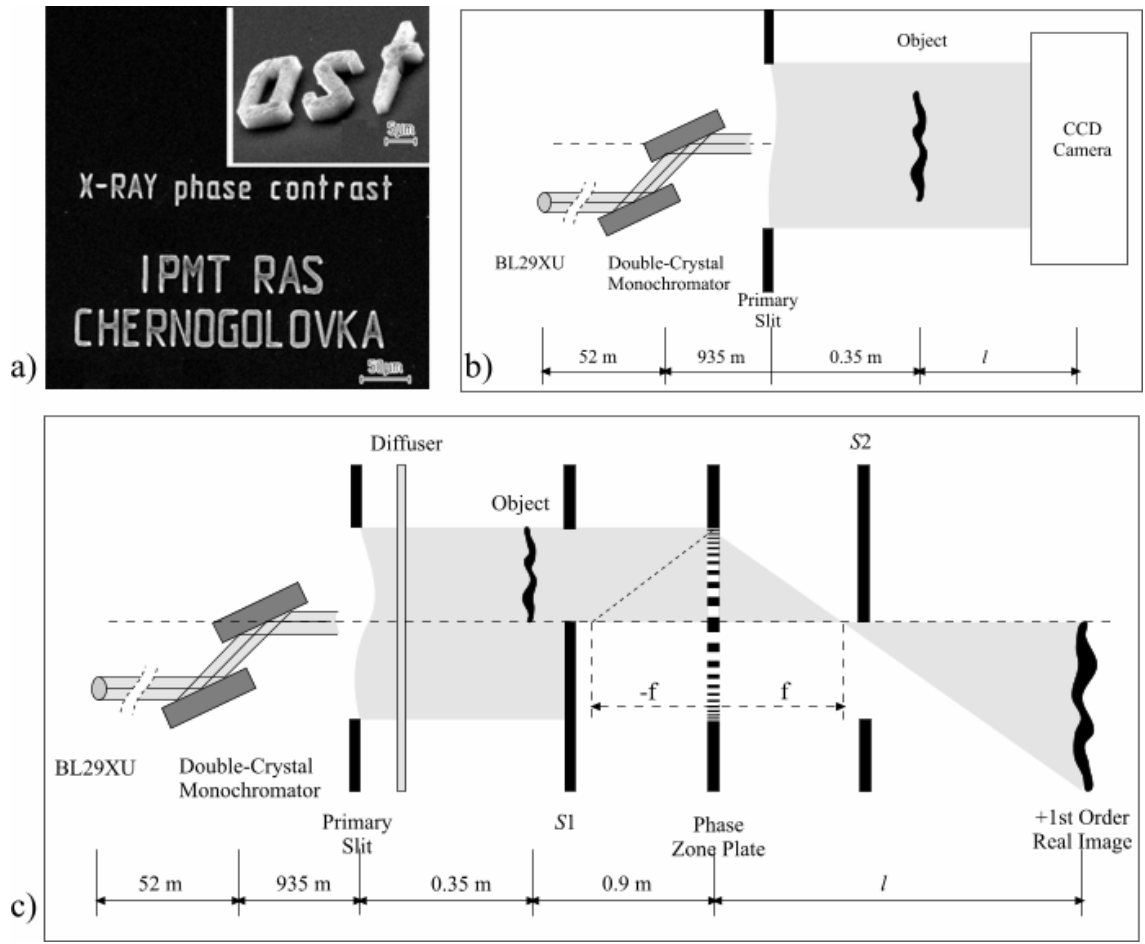

*Figure 1.* (a) SEM image of the test object. (b) Scheme of the experiment for recording a hologram image. (c) Scheme of the experiment for transmitting a hologram image.

In the second case (Fig.1c) to obtain the image reproduction the 10 KeV X-ray beam behind the test object is collimated through a 160μm x 90μm $S1$ slit and falls on the phase zone plate. Then the X-ray beam is overlapped by a 150μm x 80μm $S2$ slit. The parameters of the phase ZP used are the following: the radius of the first zone is 8.42μm, the number of zones is 112, the size of the last zone is 0.4μm, an aperture is 178.2μm, the focal distance is 57.5 cm, the relief height is 10.5μm, the membrane thickness is 16μm. The transmission coefficient of the membrane is 79% for the energy of 10keV. The average transmission coefficient of the surface relief is 93%. The reproduction of the test object image has been carried out experimentally using the order phase zone plate. The images have been recorded more than 100 times, with the step of the CCD camera equal to 1 cm. The transmission of the test object image is shown in Figures 2 (e,f,g,h) for different distances from the ZP. It should be noted that these images correspond to the fixed position of the test object, namely as far as 90 cm from the ZP.

The reproduction of the image has also been investigated when: a) the test object was situated between the ZP and the minus first order of the phase zone plate, it is an imaginary image; b) the test object was situated as far as ±3mm from the minus first order of the ZP, the image of the test object goes to infinity. In both cases the image of the test object at the different planes of observation has not been recorded.

Based on the additional results obtained it makes it possible to state that the lens formula for geometrical optics is applied also for the ZP. But on the other hand, comparing the images obtained one can see that they differ only in magnification, which is dubious for geometrical optics.

**Discussion of experimental results**

The contrast obtained cannot be of an amplitude character because the absorption coefficient of the silicon test object is comparatively small. Moreover, the amplitude contrast should be observed in the nearest field. The obtained contrast cannot be of a phase character only because the diffuser very weakly affects the image. To transmit the image by the ZP, a first-order phase plate, which gives a $\pi/2$ shift, should be installed in the focus [3- 5]. But when transmitting the image a phase plate has not been used in the scheme. Consequently the contrast obtained cannot be of a phase character only even though the 3mm (for the energy of 10keV) letters on the test object result in a $0.3\pi$ shift.

The reason for the image formation is diffraction on the letters which has an edge enhanced contrast at letters boundaries where appearing a jump of the phase. After the diffraction the beam gains a certain divergence $\alpha = \lambda/d$, where $\lambda$ is the radiation wavelength, $d$ is the letter width; then the beam is interfered by the basic beam which travels past the letters without deflection and forms a holographic image (a Gabor hologram), which showed in Fig.2a,b. When fixing the coordinates of the test object the image obtained (Fig.2c,d,e,f) is a direct evidence that the image formed is holographic. Since a three-dimensional image is continuously formed behind the object up to the ZP, the same contrast with different magnification is obtained when transmitting the image.

**Conclusion**

The reproduction of a hologram image for hard X-ray radiation has been carried out and recorded for the first time. The results of the work open prospects to create new diagnostic devices for X-ray radiation. With advances in nanotechnologies a ZP with the resolving power of several nanometers can be created, and a hologram image of an object with the magnification from $10-10^4$ times can be reproduced, which has specific application in both biology and nanotechnology.

**Acknowledgements**

The authors are grateful to Prof. V. Kohn, Dr. E. Shulakov for the assistance in discussing results.

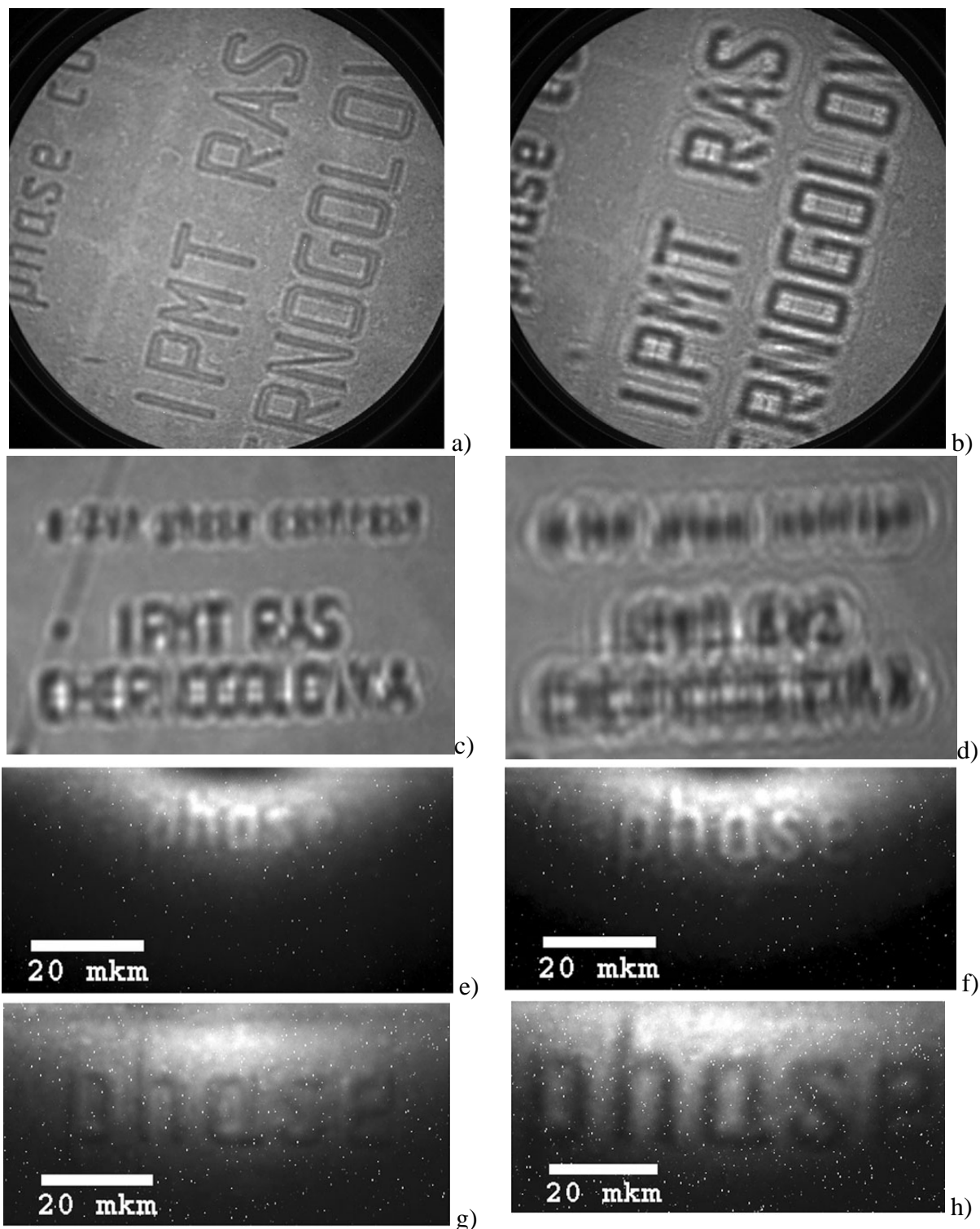

*Figure 2.* **Experimental results. The hologram image of the test object at the distance:
(a)** $l = 75$ mm, **(b)** $l = 325$ mm.
**Fourier hologram in Fraunhofer region, distance from test object to image plane: (c)
L=95 cm, (d) L=260 cm.**
**Transmission of a hologram image by a zone plate at the distance: (e)** $l=115$ cm, **(f)** $l=130$ cm, **(g)** $l=160$ cm, **(h)** $l=190$ cm.